\documentclass[aps,pra,twocolumn,showpacs,preprintnumbers,amsmath,amssymb]{revtex4-1}
\usepackage{ifpdf}
 \newif\ifpdf
\ifx\pdfoutput\undefined
   \pdffalse
\else
   \pdfoutput=1
   \pdftrue
\fi
\ifpdf
   \usepackage{graphicx}
   \usepackage{epstopdf}
   \usepackage{color}
\else
   \usepackage{graphicx}
\fi

\usepackage{srctex}
\usepackage{hyperref}
\usepackage{color}

\newcommand{ \be}{\begin{equation}}
\newcommand{ \ee}{\end{equation}}
\newcommand{\beq}{\begin{eqnarray}}
\newcommand{\eeq}{\end{eqnarray}}
\newcommand{\bem}{\begin{pmatrix}}
\newcommand{\eem}{\end{pmatrix}}
\newcommand{\bmx}{\begin{array}}
\newcommand{\emx}{\end{array}}

\DeclareMathOperator{\Imag}{Im}\DeclareMathOperator{\Real}{Re}
\begin{document}


\title{Pade spectroscopy of structural correlation functions: application to liquid gallium }

\author{N.M. Chtchelkatchev}
\affiliation{L.D. Landau Institute for Theoretical Physics, Russian Academy of Sciences, 117940 Moscow, Russia}
\affiliation{Moscow Institute of Physics and Technology, 141700 Moscow, Russia}
\affiliation{Institute for High Pressure Physics, Russian Academy of Sciences, 142190 Troitsk, Russia}

\author{B.A. Klumov}
\affiliation{High Temperature Institute, Russian Academy of Sciences, 125412, Izhorskaya 13/2, Russia}
\affiliation{L.D. Landau Institute for Theoretical Physics, Russian Academy of Sciences, 117940 Moscow, Russia}

\author{R.E. Ryltsev}
\affiliation{Institute of Metallurgy, Ural Division of Russian Academy of Sciences, Yekaterinburg 620016, Russia}
\affiliation{L.D. Landau Institute for Theoretical Physics, Russian Academy of Sciences, 117940 Moscow, Russia}
\affiliation{Ural Federal University, 620002 Yekaterinburg, Russia}

\author{R.M. Khusnutdinoff}
\affiliation{Kazan (Volga Region) Federal University, ul. Kremlevskaya Str. 18, Kazan, 420008, Russia }
\affiliation{L.D. Landau Institute for Theoretical Physics, Russian Academy of Sciences, 117940 Moscow, Russia}

\author{ A.V. Mokshin}
\affiliation{Kazan (Volga Region) Federal University, ul. Kremlevskaya Str. 18, Kazan, 420008, Russia }
\affiliation{L.D. Landau Institute for Theoretical Physics, Russian Academy of Sciences, 117940 Moscow, Russia}

\date{\today}
\begin{abstract}
We propose the new method of fluid structure investigation which is based on numerical analytical continuation of structural correlation functions with Pade approximants. The method particularly allows extracting hidden structural features of non-ordered condensed matter systems from experimental diffraction data. The method has been applied to investigating the local order of liquid gallium which has non-trivial stricture in both the liquid and solid states. Processing the correlation functions obtained from molecular dynamic simulations, we show the method proposed reveals non-trivial structural features of liquid gallium such as the spectrum of length-scales and the existence of different types of local clusters in the liquid.
\end{abstract}

\maketitle
\section{Introduction}

Gallium is a very specific metal~\cite{Waseda1980structure}. First of all, it has enormously large domain of its phase diagram  corresponding to liquid state. For example, the temperature interval of stable liquid state at ambient pressure is $(302.93, 2477)$ K. Diffraction experiments have shown that the local structure of liquid gallium is very complicated and it differs much from the local structure of typical simple liquids~\cite{Narten1972JChemPhys,Waseda1972PhysStatSol,Brazkin2008JETP,Brazhkin2012PhysRevB}. Accordingly, the crystal structure of gallium is also nontrivial: below its melting temperature $T_m = 302.93 $ at the ambient pressure the stable phase corresponds to the orthorhombic lattice which in nontypical for one-component metallic systems \cite{Wells2012StructChem} (amount the other metals, the only \textrm{Eu} has the same lattice structure at ambient pressure). At higher pressures there is a lot of polymorphic transitions to other nontrivial crystal structures~\cite{Schulte1997PRB,Kenichi1998PRB}.

Last time a lot of theoretical efforts were concentrated on investigation of the local structures and anomalies of gallium in liquid phase \cite{Tsay1994PRB,Gong1993EPL,Yang2011JChemPhys,Tsai2010JChemPhys}. But there is a gap between theoretical investigations of gallium and experiment. The main problem is that the only way to directly access the local structure is performing  molecular dynamic (MD) simulations which can not unambiguously describe structure of real materials. Indeed, classical MD deals with model approximate potentials and  first-principles MD has a problem of restricted spatial and time scales available in simulations.  Experiment also can not directly access the local structure: the information about angle correlations is mostly lost in the static structure factor or radial distribution function (RDF); so only radial correlations can be extracted.  Here we develop a new method of correlation function processing based on complex analysis with Pade approximants.

 Using this method we study local structure of liquid gallium and extract its non-trivial features. Analysing structural correlation functions obtained from MD simulations with EAM potentials, we extract the spectrum of spatial length scales and the existence of two types of local clusters. We show that the best fit of the gallium RDF can be performed using Lorentzians instead of Gaussians which are usually used for that purpose.

\begin{figure*}[t]
  \centering
  \includegraphics[width=0.99\textwidth]{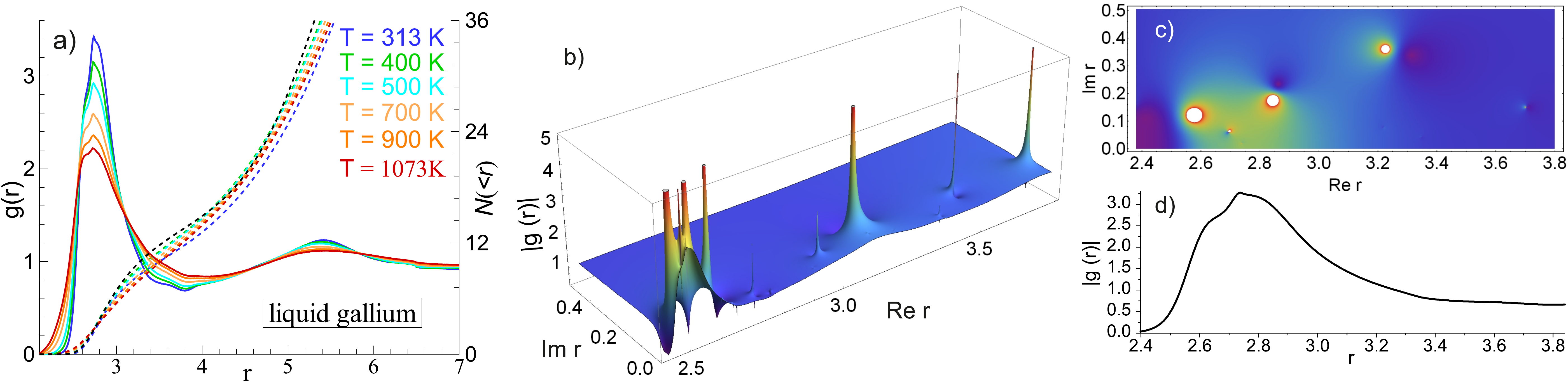}\\
  \caption{(Color online) Pade spectroscopy of radial distribution function. (a) Radial distribution function (RDF) $g(r)$ of liquid
  gallium (solid lines) and its cumulative function $N(<r)$ (dashed lines) at different temperatures $T$ (indicated on the plot).  (b)-(d) Liquid gallium, $T=500$. (b) Doing the Pade approximation we find characteristic scales of gallium for particle distribution  in the first coordination sphere:  $r=2.58$, $r=2.69$, $r=2.84$. The scale $r=3.33$ corresponds to the tail of the first coordination sphere. (c) shows the density plot of the Pade approximant while (d) shows for comparison the first peak of $g(r)$.} \label{figrdfs}
\end{figure*}


\section{Local order of gallium}
\subsection{Methods}
For MD simulations of liquid gallium, we have used $\rm{LAMMPS}$ molecular dynamics package. The system of $N=20000-100000$ particles interacting via EAM potential~\cite{Belashchenko2013UFN}, specially designed for gallium \cite{Belashchenko2012JPhysChemA}, was simulated under periodic boundary conditions in Nose-Hoover  NPT ensemble. This amount of particles is enough to obtain satisfactory results.  More simulation details can be found in Ref.~\cite{Mokshin2015JETP} where similar simulation has been described. We checked that radial distribution functions obtained by our simulation quantitatively agree with those extracted from experiment~\cite{Mokshin2015JETP}.

\subsection{Local order: Pade spectroscopy of pair  radial distribution}

\begin{figure}[t]
  \centering
  \includegraphics[width=0.9\columnwidth]{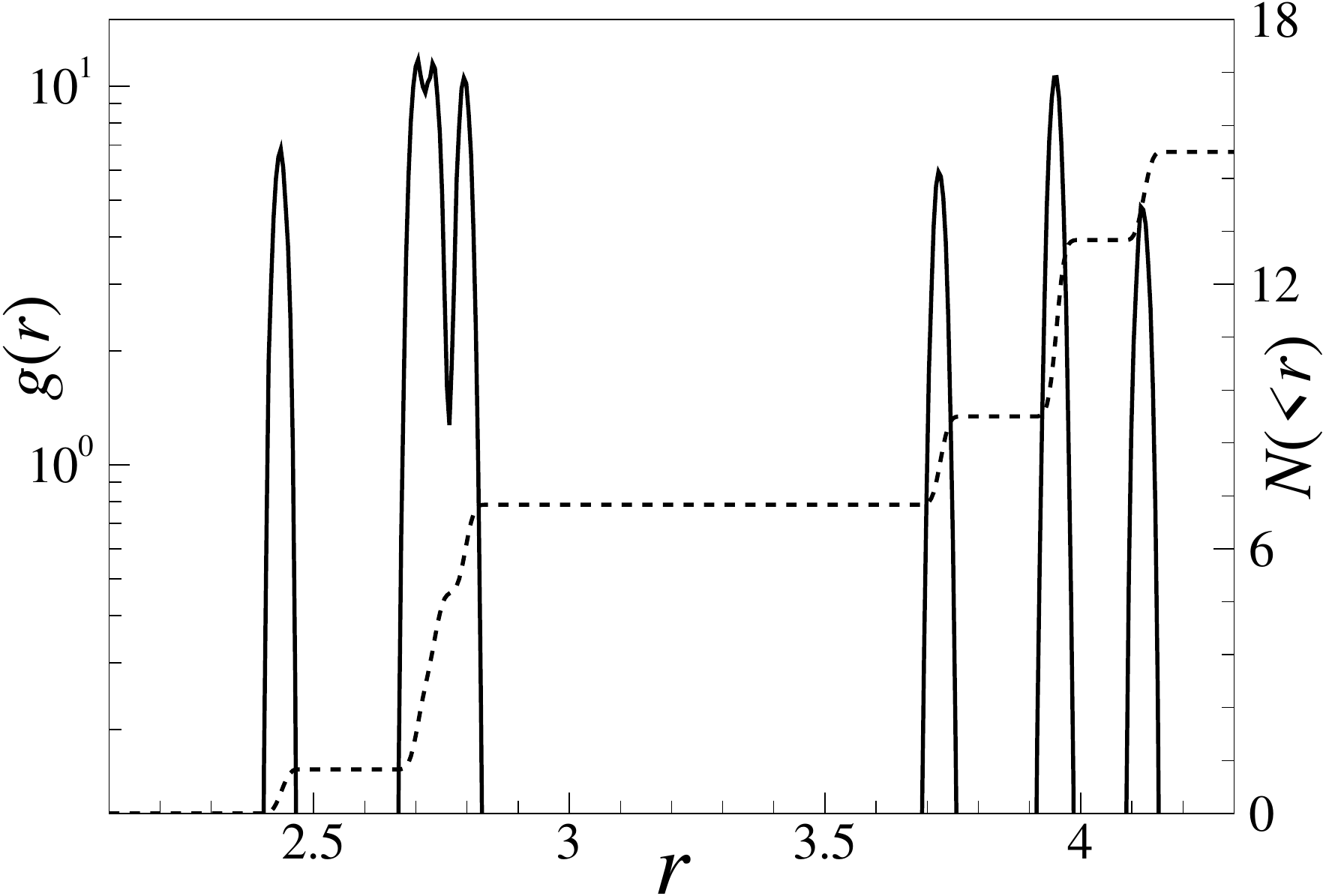}\\
  \caption{(Color online) RDF obtained for ideal lattice of $\alpha$-\textrm{Ga} with small random noise which mimics thermal vibrations. The cumulative function $N(<r)$ is also plotted by dashed line to evaluate the number of atoms at different shells. }\label{rdf_cryst}
\end{figure}
Here we focus on gallium at ambient pressure and investigate how its properties in liquid state depend on temperature. Figure~\ref{figrdfs}(a) shows the radial distribution function $g(r)$ of gallium taken for several temperatures in the range of $(313, 1073)$ K. The cumulative function $N(<r) \equiv 4\pi \rho \int_0^r r'^2 g(r') dr'$, which is the mean number of particles inside a sphere of radius $r$, is also plotted in Fig.~\ref{figrdfs}(a); in part, the cumulative function $N(<r)$  shows the number of nearest neighbors in the first coordination shell.

It is clear from the figure that the shape of the first RDF peak is nontrivial.  It can be seen that the first RDF peak of gallium has clear shoulder.
Particles forming the local clusters are mostly located at distances corresponding to the first coordination sphere (first peak of RDF). That suggests that local structure of liquid gallium is rather nontrivial. More detailed information can be extracted only after specific processing of RDF (see below).

The promising way to extract features of local order hidden in RDF is performing analytical continuation to complex plain of distances. This is usual way in the theory of correlation functions of quantum systems especially for analytical continuation of the Greens functions from imaginary (Matsubara~\cite{Abrikosov2009AGDBook}) frequencies to the real frequency domain~\cite{EliashbergEqPade,yamada2013analyticity,schott2015arXiv}. Recently the method of numerical analytical continuation was successfully applied to investigations of velocity autocorrelation function of Lennard-Jones fluid~\cite{Krishnan2003JCP,Chtchelkatchev2015JETPLett} and in classical hydrodynamics of the Stokes waves~\cite{dyachenko2014JETPLett}. Here we use this method, see Appendix~\ref{ApPade} and Ref.~\cite{Chtchelkatchev2015JETPLett}, for analysis of gallium RDFs.

\begin{figure*}[t]
  \centering
  \includegraphics[width=0.99\textwidth]{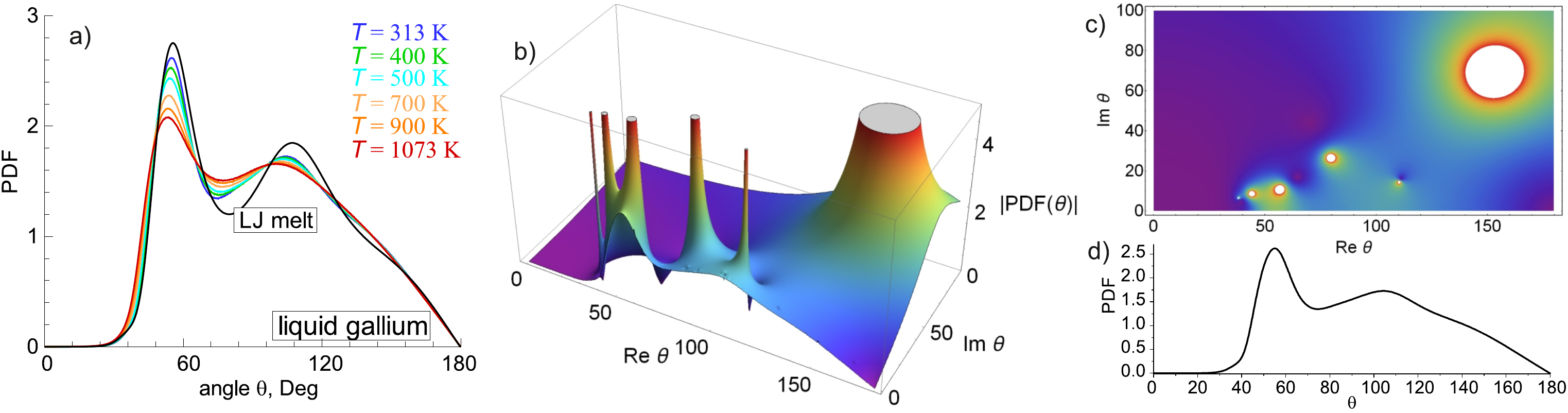}\\
  \caption{(Color online) Pade spectroscopy of bond angle distribution function (BADF). (a) BADF of liquid gallium at different temperatures $T$ (indicated on the plot). Additionally the BADF of LJ
  melt (which is nearly universal on the LJ melting curve) is also plotted for the comparison. (b-d) show BADF for $T=500$ and its processing with Pade approximant. We see a number of poles, their positions ($\Real \theta$) give characteristic angles between the particles forming local order clusters. }\label{figpdfas}
\end{figure*}
\begin{figure*}[t]
  \centering
  \includegraphics[width=0.99\textwidth]{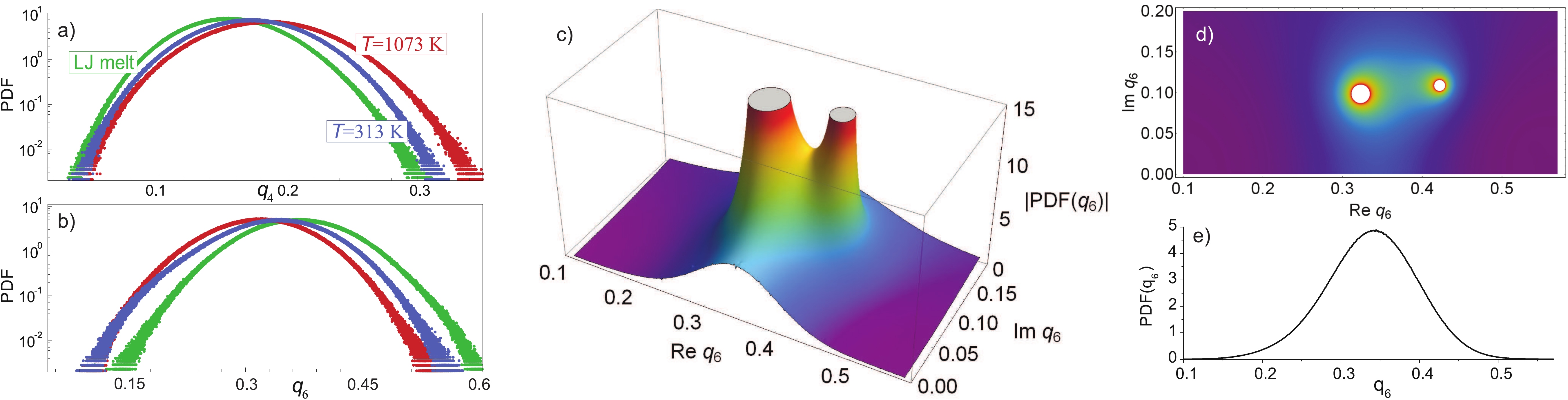}\\
  \caption{(Color online) Pade spectroscopy of probability distribution functions (PDF). (a) and (b) show orientational local order of liquid gallium: PDFs of the bond order parameters $q_l$ ($l$ = 4, 6) taken at two temperatures $T \--$ 313 K (blue line) and 1073 K (red line). Green solid lines represent the same PDFs for the LJ melt taken on the melting line (it can be shown that along the melting line these PDFs are practically universal \cite{KlumovPrivateComm}). (c) and (d) show the absolute value of PDF ($l=6$) of gallium for $T=500$ in complex $q_6$-plain. (e) is PDF of gallium for $T=500$ plotted for comparison with poles of the PADE-approximant in (d). The poles are situated at $\Real q_6=0.32,\,0.42$.}\label{figpdfpade}
\end{figure*}
The static structure factor is related to RDF as~\cite{hansenMcDonald}:
\begin{gather}
 S(k)=1+\frac {4\pi}\rho\Imag\int_0^\infty r (g(r)-1)e^{ikr}dr.
\end{gather}
So, if $g(r)$ has a pole, at the position $r_p$, we should expect that the contribution of this pole to the structure factor oscillates like $\cos(k\Real r_p)$ or $\sin(k\Real r_p)$  and decays with $k$ like $e^{-k\Imag r_p}$. Thus, knowledge of the poles gives important scales characterising the particle system.

In Figs.~\ref{figrdfs}(b) and (c) we show the results of the typical processing of gallium RDF by Pade approximants. As follows we use Pade approximant of RDF for its analytical continuation in complex-$r$. Fig.~\ref{figrdfs}(b) shows the absolute value of RDF in complex-$r$ plain while the peaks correspond to the poles. We see a limited number of poles near the real axis: so the Lorenzian fit of RDF perfectly matches all its basic features, see  Figs.~\ref{figrdfs}(c)  and (d).  The real parts of the positions of the poles are important length scales characterising gallium: as follows, in the first coordination sphere:  $r=2.58$, $r=2.69$, $r=2.84$;  the scale $r=3.33$ corresponds to the tail of the first coordination sphere. These scales are in fact characteristic interparticle distances in the first coordination shell. Thus we see that analytical continuation with Pade approximants allows extracting multi-scale character of local structure of liquid gallium. In that connection, remember that, at ambient pressure, gallium crystallizes into non-trivial orthorhombic lattice and so it is interesting to find relation between unusual multiscale local order of the liquid and non-trivial crystal symmetry.

In Fig.\ref{rdf_cryst} we show RDF obtained for ideal lattice of $\alpha$-\textrm{Ga} (orthorhombic) with small random noise which mimics thermal vibrations. We see that first coordination shell contains two atoms located at short distance of 2.46 ${\AA}$. Such ${\rm Ga_2}$ dimers were earlier suggested to be formed by covalent-like bonding \cite{Gong1991PRB}. Each Ga atom has another six nearest neighbour located in pairs in different coordination shells at distances $2.7$, $2.73$, $2.79$ ${\AA}$. So the crystal structure of $\alpha$-{\rm Ga} demonstrates spectrum of spatial scales. The comparison of these scales with those obtained from analytical continuation of RDF shows good agreement. Indeed the ratios of maximal and minimal distances are $1.134$ for experimental lattice structure and $1.1$ for scales extracted from liquid RDF. Of course the analytical continuation of liquid RDF does not distinct two nearly located scales. Recently, the similar analysis of liquid Ga  structure was performed in \cite{Mokshin2015JETP} where simulated RDF were approximated by a set of Gaussians that gave similar interval of spatial scales. However this methods can distinct only two scales in mentioned interval. Moreover the results strongly depend on the choice of approximation parameters such as the number and the shape of approximating functions.




\subsection{Local order: Pade spectroscopy of orientational order}

More detailed structural information can be obtained from analysis of orientational local order. The simplest way to describe it is calculating angular distribution function $P(\theta)$ which is probability density of angle $\theta$ between two vectors connecting a particle with its two nearest neighbors.

 Fig.~\ref{figpdfas} shows bond angle distribution functions of liquid gallium at different temperatures $T$ (indicated on the plot). Additionally the BADF of LJ-melt (which is nearly universal on the LJ melting curve) is also plotted for the comparison. Figure~\ref{figpdfas}(b-d) show BADF for $T=500$. Looking at BADFs at real-$\theta$ axis, it is difficult to say something specific about the clusters forming the local order.   But after its processing with the Pade approximant we see a number of poles in complex-$\theta$ plain. The positions of poles ($\Real \theta$) give characteristic angles between the particles forming the local order clusters. Particulary, we see two pronounced poles at the vicinity of the first peak which are located near the angles $45$ and $60$. The value of $\theta=60$ is typical for simple liquids and corresponds to tetrahedral local order \cite{Ryltsev2013PRE}. But the angle $\theta=45$ suggests non-trivial symmetry of local clusters probably caused by the existence of short-bonded particles revealed by RDF analysis.


Another way to analyse orientational order is the well known bond order parameter method, which is widely used to characterize order in simple fluids, solids and glasses~\cite{Steinhardt1983PRB,Steinhardt1981PRL,Errington2003JChemPhys,TenWolde,KlumovLJa,Fomin14} hard spheres~\cite{RintoulJCP, KlumovHS}, colloidal suspensions~\cite{Gasser01}, complex plasmas~\cite{RubinZ,KlumovPU, Khrapak2011} and metallic glasses \cite{Hirata2013Science}.

Basic description of bond order distribution functions (BODF) can be found in Appendix~\ref{ApBond}.
To quantify the local orientational order, it is also convenient to use the probability distribution functions (PDFs) $P(q_l)$ and $P(w_l)$. Figure~\ref{figpdfpade} shows
the PDFs at different $l$ ($l=4,~6$) at different temperatures of liquid gallium in comparison to those calculated for LJ liquid whose PDFs are nearly universal along the melting line~\cite{KlumovPrivateComm}. We se again that such PDFs calculated at real-$q_l$ axis (Fig.~\ref{figpdfpade}a,b) reveal no interesting features of liquid gallium structure. We see broad dome-shaped distributions which are similar to those for LJ liquids. But analytical continuation in the complex-$q$ plane reveals that $P(q_l)$ are in fact composed of two Lorenzian-like peaks with similar values of both the maximum location (${\rm Re}(q)$) and peak width (${\rm Im}(q)$). This fact suggests the structure of liquid gallium consists of two types of local order that is in agreement with the earlier obtained results \cite{Brazhkin2012PhysRevB,Yang2011JChemPhys}.




\section{Discussion}
\begin{figure}[t]
  \centering
  \includegraphics[width=0.99\columnwidth]{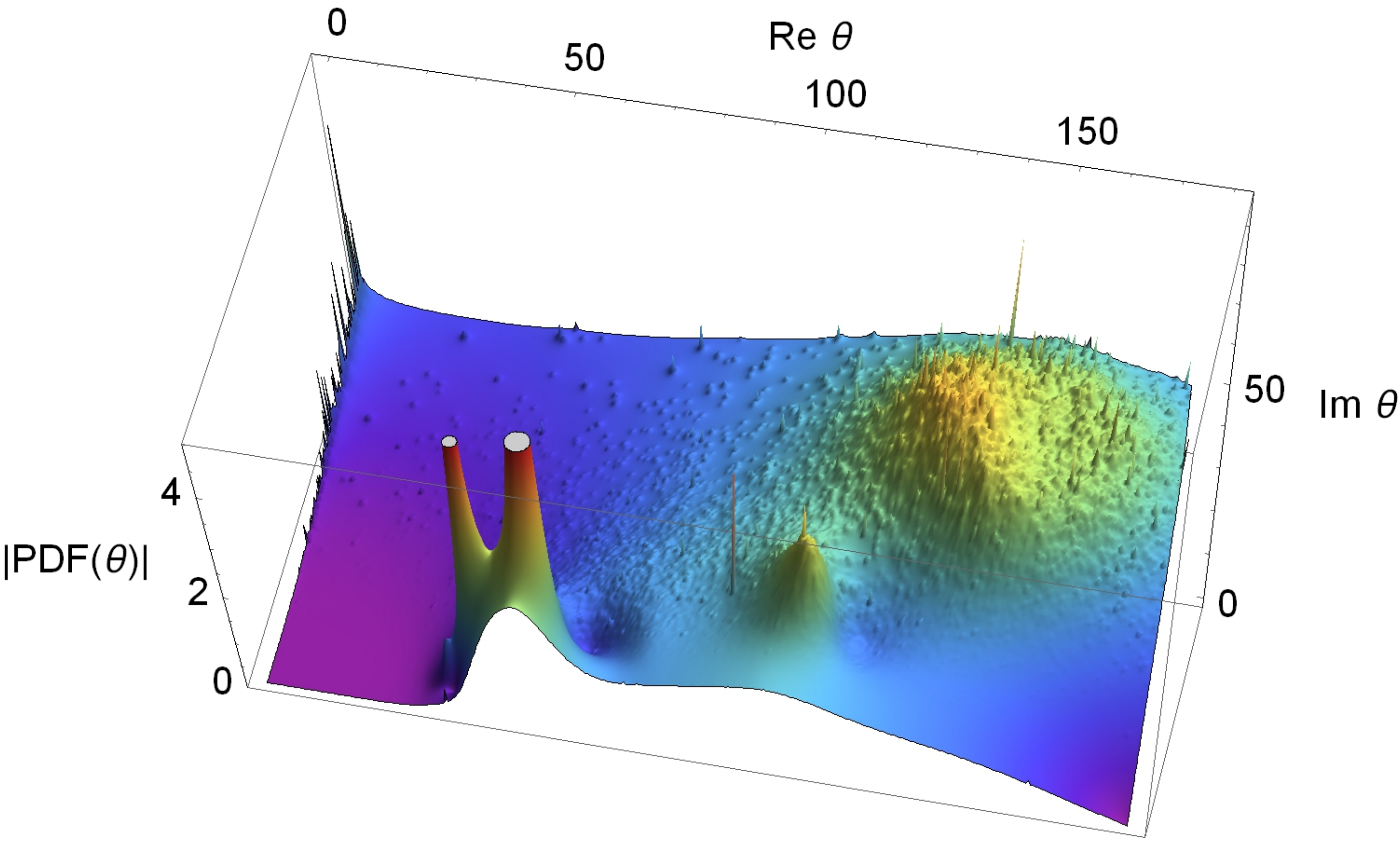}\\
  \caption{(Color online) Stability test of PDF analytical continuation shown in Fig.~\ref{figpdfas}(b).}\label{PDF_noise}
\end{figure}
There is important question related to the accuracy of numerical analytical continuation and stability of poles to errors in the initial data. We solve this problem doing two procedures.

We build the Pade-approximant on top known function values $f(x_i)=u_i$ in the discrete set of points $x_i$ on the real axis, see Appendix~\ref{ApPade}. We always randomly shuffle the set of points $x_i$ using the Fisher--Yates--Knuth shuffle algorithm~\cite{Knuth1997} before building the Pade  continued fraction approximant. Then we average the approximant over a considerable number of shuffles. This procedure helps avoiding specific errors related to numerical calculation of long continued-fraction.

Next, the accuracy of the Pade extrapolation of the function depends on the distribution of points $x_i$: maximum accuracy corresponds to the uniform distribution of $x_i$. Doing the stability test we randomly take away about 10$\%$ of points $x_i$: this procedure makes the distribution of $x_i$ nonuniform and moves the ``unstable'' poles in the complex plain if there are any. Finally we build the extrapolation $\tilde f$ of the function $f$ averaging the Pade approximant over $10000$ random extractions of initial points $x_i$. Then when we draw $|\tilde f(z)|$ at complex $z$, unstable poles disappear while the poles suffering from the insufficient accuracy of $f(x_i)$ average into ``domes''.

This procedure is illustrated in Fig.~\ref{PDF_noise}. We test here PDF shown in Fig.~\ref{figpdfas}(b) using the procedure described above. Poles at $\Real\theta\approx 44$ and $\Real\theta\approx 57$ are stable. The other poles shown in  Fig.~\ref{figpdfas}(b) are not wrong, but their positions are more sensitive to the accuracy of the PDF-table. So after each random taking away some of the knots $\theta_i$, the positions of the poles slightly change and finally after 10000 such procedures we see domes at the positions where in  Fig.~\ref{figpdfas}(b)  we have seen poles. This test show that we can trust all the pole except, probably, one situated at $\Real \theta\approx79$. (More refined stability test however shows that even this pole should not be completely disregarded.)

This procedure is illustrated in Fig.~\ref{PDF_noise}. We test here PDF shown in Fig.~\ref{figpdfas}(b) using the method described above. Poles at $\Real\theta\approx 44$ and $\Real\theta\approx 57$ are stable. The other poles shown in  Fig.~\ref{figpdfas}(b) are not wrong, but their positions are more sensitive to the accuracy of the PDF-table. So after each random taking away some of the knots $\theta_i$, the positions of the poles slightly change and finally after averaging over 10000 such extractions we see domes at the positions where in  Fig.~\ref{figpdfas}(b)  we have seen poles. This test shows that we can trust all the poles except, probably, one situated at $\Real \theta\approx79$. (More refined stability test however shows that even this pole should not be completely disregarded.) The coordinates of the ``dome'' top in the complex plain we should take then as the characteristic parameters. If we compare them with the coordinates of the poles in Fig.~\ref{figpdfas}(b) we will see coincidence with the satisfactory accuracy.

\section{Conclusions}

We propose the new method of fluid structure investigation which is based on numerical analytical continuation of structural data obtained from both experiment and computer simulations. The method particularly allows extracting hidden structural features of non-ordered condensed matter systems from experimental diffraction data. The method has been applied to investigating the local order of liquid gallium which is supposed to have complex stricture. We show that analytical continuation of structural correlation functions such as radial distribution function, bond angle distribution function and bond orientational order parameters reveals non-trivial structural features of liquid. Firstly, we show that, processing the liquid RDF, our method allows easily obtaining the spectrum of length-scales which are in close agreement with those for crystal state. Secondly, we show for the first time that correlation functions of orientational order also have non-trivial features probably caused by the existence of different types of local clusters in the liquid.

We also show that the best fit of the radial distribution function can be performed using  Lorentzians instead of Gaussians which are usually used for that purpose.

\acknowledgments
We thank V.  Brazhkin and  D. Belashchenko for stimulating discussions. This work was supported by Russian Science Foundation (grant №14-12-01185). We are grateful to Russian Academy of Sciences for the access to JSCC and ``Uran'' clusters.

\appendix
\section{Pade approximants method\label{ApPade}}

Here we discuss the construction of the Pad\'{e} approximants that interpolate a function given $N$ knot points.  Pade-approximants are the rational functions  (ratio of two polinomials).  A rational function can be represented by a continued fraction. Typically the continued fraction expansion for a given function approximates the function better than its series expansion. Here we use ``multipoint'' algorithm~\cite{EliashbergEqPade,baker1996pade,wolframPade} to build the  Pade-approximant. Suppose we know function values $f(x_i)=u_i$ in the discrete set of points $x_i$ where the ``knots'' $x_i$, $i=1,2,3,\ldots,N$. Then the Pade approximant
\begin{gather}
C_N(x)=\frac{a_1}{\frac{a_2\left(x-x_1\right)}{\frac{a_3 \left(x-x_2\right)}{\frac{a_4\left(x-x_3\right)}{\ldots  a_N\left(x-x_{N-1}\right)+1}+1}+1}+1}
\end{gather}
where  $a_i$ we determine using the condition, $C_N(x_i)=u_i$, which is fulfilled if $a_i$ satisfy the recursion relation
\begin{gather}\label{eq2}
  a_i=g_i\left(x_i\right), \qquad
g_1\left(x_i\right)=u_i, \qquad i=1,2,3,\ldots ,N.
\\\label{eq3}
g_p(x)=\frac{g_{p-1} (x_{p-1})-g_{p-1}(x)}{\left(x-x_{p-1}\right) g_{p-1}(x)},\qquad p\geq 2.
\end{gather}
Eq.~\eqref{eq2} is the ``boundary condition'' for the recursive relation Eq.~\eqref{eq3}. For example, taking $x=x_{i_0}$ we get $g_1(x_{i})$ from~\eqref{eq2} and $g_j(x_{i_0})$, $j=2,3,\ldots,i$ from~\eqref{eq3}. Accuracy of the analytical continuation can be tested by permutations of  $x_i$ and/or by the random extraction of some part of the knots and permutation averaging.

\section{Bond orientational order analysis\label{ApBond}}
\begin{table}[th]
\centering
\caption{Bond order parameters $q_l$ and $w_l$ ($l=4,~6$) calculated via the fixed number of nearest neighbors (NN) for a few crystalline structures and cumulants of the corresponding BODF for the LJ melt}.
\begin{tabular}{|c|c|c|c|c|}
\hline 
lattice type & \quad $q_{4}$ & \quad $q_{6}$ & \quad $w_{4}$ & \quad $ w_{6}$
\\ \hline hcp (12 NN) & 0.097 & 0.485 & 0.134  & -0.012
\\ \hline fcc  (12 NN) & 0.19  & 0.575  & -0.159 &  -0.013
\\ \hline ico  (12 NN) & $1.4 \times 10^{-4}$ & 0.663 & -0.159  & -0.169
\\ \hline bcc ( 8 NN) & 0.5 & 0.628 & -0.159   & 0.013
\\ \hline bcc (14 NN) & 0.036 & 0.51 & 0.159   & 0.013
\\ \hline LJ melt (12 NN) & $\approx $0.155  & $\approx $0.37  & $\approx $-0.023  &$\approx $-0.04
\\\hline 
\end{tabular}
\label{t1}
\end{table}

Each particle $i$ is connected via vectors (bonds) with its $N_{\rm nn}(i)$ nearest neighbors (NN), and the rotational invariants (RIs) of rank $l$ of second $q_l(i)$ and third $w_l(i)$ orders are calculated as:
\be
q_l(i) = \left ( {4 \pi \over (2l+1)} \sum_{m=-l}^{m=l} \vert~q_{lm}(i)\vert^{2}\right )^{1/2}
\ee
\be
w_l(i) = \hspace{-0.8cm} \sum\limits_{\bmx {cc} _{m_1,m_2,m_3} \\_{ m_1+m_2+m_3=0} \emx} \hspace{-0.8cm} \left [ \bmx {ccc} l&l&l \\
m_1&m_2&m_3 \emx \right] q_{lm_1}(i) q_{lm_2}(i) q_{lm_3}(i),
\label{wig}
\ee
\noindent
where $q_{lm}(i) = N_{\rm nn}(i)^{-1} \sum_{j=1}^{N_{\rm nn}(i)} Y_{lm}({\bf r}_{ij} )$, $Y_{lm}$ are the spherical harmonics and ${\bf r}_{ij} = {\bf r}_i - {\bf r}_j$ are vectors connecting centers of particles $i$ and $j$. In Eq.(\ref{wig}), $\left [ \bmx {ccc} l&l&l \\ m_1&m_2&m_3 \emx \right ]$ are the Wigner 3$j$-symbols, and the summations performed over all the indexes $m_i =-l,...,l$ satisfying the condition $m_1+m_2+m_3=0$. As shown in the pioneer paper\cite{Steinhardt1983PRB}, the bond order parameters $q_l$ and $w_l$ can be used as measures to characterize the local orientational order and the phase state of considered systems.

Because each lattice type has a unique set of bond order parameters, the method of RIs can be also used to identify lattice and liquid structures in mixed systems. The values of $q_l$ and $w_l$ for a few common lattice types (including liquid-like
Lennard-Jones melt) are presented in Table~\ref{t1}.

 \bibliography{our_bib_corr_n}
\end{document}